# MMV_Im2Im: An Open Source Microscopy Machine Vision Toolbox for Image-to-Image Transformation

## Authors

- **Justin Sonneck**
  [0000-0002-1640-3045](#) · [Justin-Sonneck](#) · [JustinSonneck](#)
  Leibniz-Institut für Analytische Wissenschaften - ISAS - e.V., Dortmund 44139, Germany

- **Jianxu Chen**
  [0000-0002-8500-1357](#) · [jxchen01](#) · [JianxuChen](#)
  Leibniz-Institut für Analytische Wissenschaften - ISAS - e.V., Dortmund 44139, Germany

## Abstract

Over the past decade, deep learning (DL) research in computer vision has been growing rapidly, with many advances in DL-based image analysis methods for biomedical problems. In this work, we introduce MMV_Im2Im, a new open-source python package for image-to-image transformation in bioimaging applications. MMV_Im2Im is designed with a generic image-to-image transformation framework that can be used for a wide range of tasks, including semantic segmentation, instance segmentation, image restoration, and image generation, etc.. Our implementation takes advantage of state-of-the-art machine learning engineering techniques, allowing researchers to focus on their research without worrying about engineering details. We demonstrate the effectiveness of MMV_Im2Im on more than ten different biomedical problems, showcasing its general potentials and applicabilities.

For computational biomedical researchers, MMV_Im2Im provides a starting point for developing new biomedical image analysis or machine learning algorithms, where they can either reuse the code in this package or fork and extend this package to facilitate the development of new methods. Experimental biomedical researchers can benefit from this work by gaining a comprehensive view of the image-to-image transformation concept through diversified examples and use cases. We hope this work can give the community inspirations on how DL-based image-to-image transformation can be integrated into the assay development process, enabling new biomedical studies that cannot be done only with traditional experimental assays. To help researchers get started, we have provided source code, documentation, and tutorials for MMV_Im2Im at https://github.com/MMV-Lab/mmv_im2im under MIT license.

# Introduction

With the rapid advancements in the fields of machine learning (ML) and computer vision, computers can now transform images into new forms, enabling better visualization [1], better animation [2] and better information extraction [3] with unprecedented and continuously growing accuracy and efficiency compared to conventional digital image processing. These techniques have recently been adapted for bioimaging applications and have revolutionized image-based biomedical research [4,5,6,7]. In principal, these techniques and applications can be formulated as a general image-to-image transformation problem, as depicted in the central panel in Figure 1. Deep neural networks are trained to perceive the information from the source image(s) and reconstruct the learned knowledge from source images(s) in the form of a new image(s) of the target type. The source and target images can be real or simulated microscopy images, segmentation masks, or their combinations, as exemplified in Figure 1. Since these underlying methods share the same essential spirit, a natural question arises: is it possible to develop a single generic codebase for deep learning (DL) based image-to-image transformation applicable to various biomedical studies?

In this paper, we introduce *MMV_Im2Im* an open-source microscopy machine vision (MMV) toolbox for image-to-image transformation and demonstrate its applications in over 10 biomedical applications of various types. Currently, *MMV_Im2Im* supports 2D~5D microscopy images for supervised image-to-image translation (e.g., labelfree determination [4], imaging modality transformation [5,8]), supervised image restoration [6], supervised semantic segmentation [9], supervised instance segmentation [10,11], unsupervised semantic segmentation [12], unsupervised image to image translation and synthetization [13]. The toolbox will continuously grow with more and more methods, such as self-supervised learning based methods, ideally also with contributions from the open-source community.

Why do we need such a single generic codebase for all deep-learning based microscopy image-to-image transformation? *MMV_Im2Im* is not simply a collection of many existing methods, but with rather systematic design for generality, flexibility, simplicity and reusability, attempting to address several fundamental bottlenecks for image-to-image transformation in biomedical applications, as highlighted below.

## Feature 1: universal boilerplate with state-of-the-art ML engineering:

*Bottleneck: existing codd not easy to understand or to extend or re-use*.

Our package *MMV_Im2Im* employs pytorch-lightning [14] as the core in the backend, which offers numerous benefits, such as readability, flexibility, simplicity and reusability. First of all, have you ever had the moment when you wanted to extend someone's open-source code to suit your special ML needs, but found it so difficult to figure out where and how to extend, especially for complex methods? Or, have you ever encountered the situation where you want to compare the methods and code from two different papers, even solving the same problem, e.g. semantic segmentation, but not quite easy to grasp quickly since the two repositories are implemented in very different ways? It is not rare that even different researchers from the same group may implement similar methods in very different manners. This is not only a barrier for other people to learn and re-use the open-source code, but also poses challenges for developers in maintenance, further development, and interoperability among different packages. We follow the pytorch-lightning framework and carefully design a universal boilerplate for image-to-image transformation for biomedical applications, where the implementation of all the methods share the same modularized code structure. This greatly lowers the learning curve for people to read and understand the code, and makes implementing new methods or extending existing methods simple and fast, at least from an engineering perspective.

Moreover, as ML scientists, have you ever overwhelmed by different training tricks for different methods or been curious about if certain state-of-the-art training methods can boost the performance of existing models? With the pytorch-lightning backend, *MMV_Im2Im* allows you to enjoy different state-of-the-art ML engineering techniques without changing the code, e.g., stochastic weight averaging [15], single precision training, automatic batch size determination, different optimizers, different learning rate schedulers, easy deployment on different devices, distributed training on multi-GPU (even multi-node), logging with common loggers such as Tensorboard, etc.. In short, with the pytorch-lightning based universal boilerplate, bioimaging researchers can really focus on research and develop novel methods for their biomedical applications, without worrying about the ML engineering works (which are usually lack in non-computer-science labs).

## Feature 2: modularization and human-readable configuration system:

*Bottleneck: Dilemma between simplicity and flexibility*

The toolbox is designed for both people with or without extensive experience with ML and Python programming. It is not rare to find biomedical image analysis software that are very easy to use on a set of problems, but very hard to extend or adjust to other different but essentially related problems, or find some with great flexibility with tubable knobs at all levels, but unfortunately not easy for unexperienced users. To address this issue, we design the toolbox in a systematically modularized way with various levels of configurability. One can use the toolbox with a single command as simple as `run_im2im --config train_semanticseg_3d --data.data_path /path/to/data` or make customization on details directly from a human-readable configuration file, such as choosing batch normalization or instance normalization in certain layers of the model, or adding extra data augmentation steps, etc.. We provide an extensive list of more than 20 example configurations for various appplications and comprehensive documentation to address common questions for users as reference. For users preferring graphical interface, another napari plugin for MMV toolbox has been planned as the extension of *MMV_Im2Im* (see Discussion for details).

In addition, the modularization and configuration system is designed to allow not only configuring with the elements offered by the package itself, but also any compatible elements from a third-party package or from a public repository on Github. For example, one can easily switch the 3D neural network in the original *Embedseg* method to any customized U-Net from FastAI by specifying the network as `fastai.vision.models.unet`. Such painless extendability releases the power of the toolbox, amplifies the benefit of the open-source ML community and upholds our philosophy of open science.

## Feature 3: customization for biomedical imaging applications:

*Bottleneck: Not enough consideration for specific chanllengs in microscopy images in general DL toolboxes*

The original idea of a general toolbox actually stememd from the OpenMMLab project (https://openmmlab.com/), which provides generic codebases for a wide range of computer vision research topics. For instance, *MMSegmentation* (https://github.com/open-mmlab/mmsegmentation) is an open source toolbox for semantic segmentation, supporting unified benchmarking and state-of-the-art models ready to use out-of-box. It has become one of most widely used codebase for research in semantic segmentation (2K forks and 5.4K stars on Github as of March 12, 2023). This inspires us to develop *MMV_Im2Im* to facillitate research in image-to-image transformation with special focus on biomedical applications.

First of all, different from general computer vision datasets, such as ImageNet [16], where the images are usually small 2D RGB images (e.g., 3 x 256 x 256 pixels), biomedical applications usually involves large-scale high dimensional data (e.g., 500 images of 4 x 128 x 2048 x 2048 voxels). To deal with this issue, we employ the PersistentDataset in MONAI [17] with partial loading and sampling support, as well as delayed image reading powered by aicsimageio [18] as default (configurable if another dataloader is preferred). As a result, in our stress test, training a 3D nuclei instance segmentation model with more than 125,000 3D images can be conducted efficiently in a day, even with limited resource.

Second, because microscopy data is not restricted to 2D, we re-implement common frameworks, such as fully convolutional networks (FCN), conditional generative models, cycle-consistent generative models, etc., in a generic way to easily switch between different dimensionalities for training. During inference, up to 5D images (channel x time x Z x Y x X) can be directly loaded as the input without pre-splitting into smaller 2D/3D chunks.

Third, the toolbox pre-packs common functionalities specific to microscopy images. For example, we incorporate the special image normalization method introduced in [4], where only the middle chunk along Z dimension of 3D microscopy images will be used for calculating the mean and standard deviation of image intensity for standard normalization. Also, 3D light microscopy images are usually anisotropic, i.e., much lower resolution along Z than XY dimension. So, we adopt the anisotropic variation of UNet as proposed in [19].

Finally, to deploy the model in production, a model trained on small 3D patches sometimes need to be applied not only on much large images. Combining the efficient data handling of aicsimageio [18] and the sliding window inference with gaussian weighted blending, the toolbox can yield efficient inference without visible stitching artifacts in production.

All in all, the *MMV_Im2Im* toolbox stands on the shoulders of many giants in the open-source software and ML engineering communities (pytorch-lightning, MONAI, aicsimageio, etc.) and is systematically designed for image-to-image transformation R&D for biomedical applications. The source code of *MMV_Im2Im* is available at https://github.com/MMV-Lab/mmv_im2im. This manuscript is generated with open-source package Manubot [20]. The manuscript source code is available at https://github.com/MMV-Lab/im2im-paper.

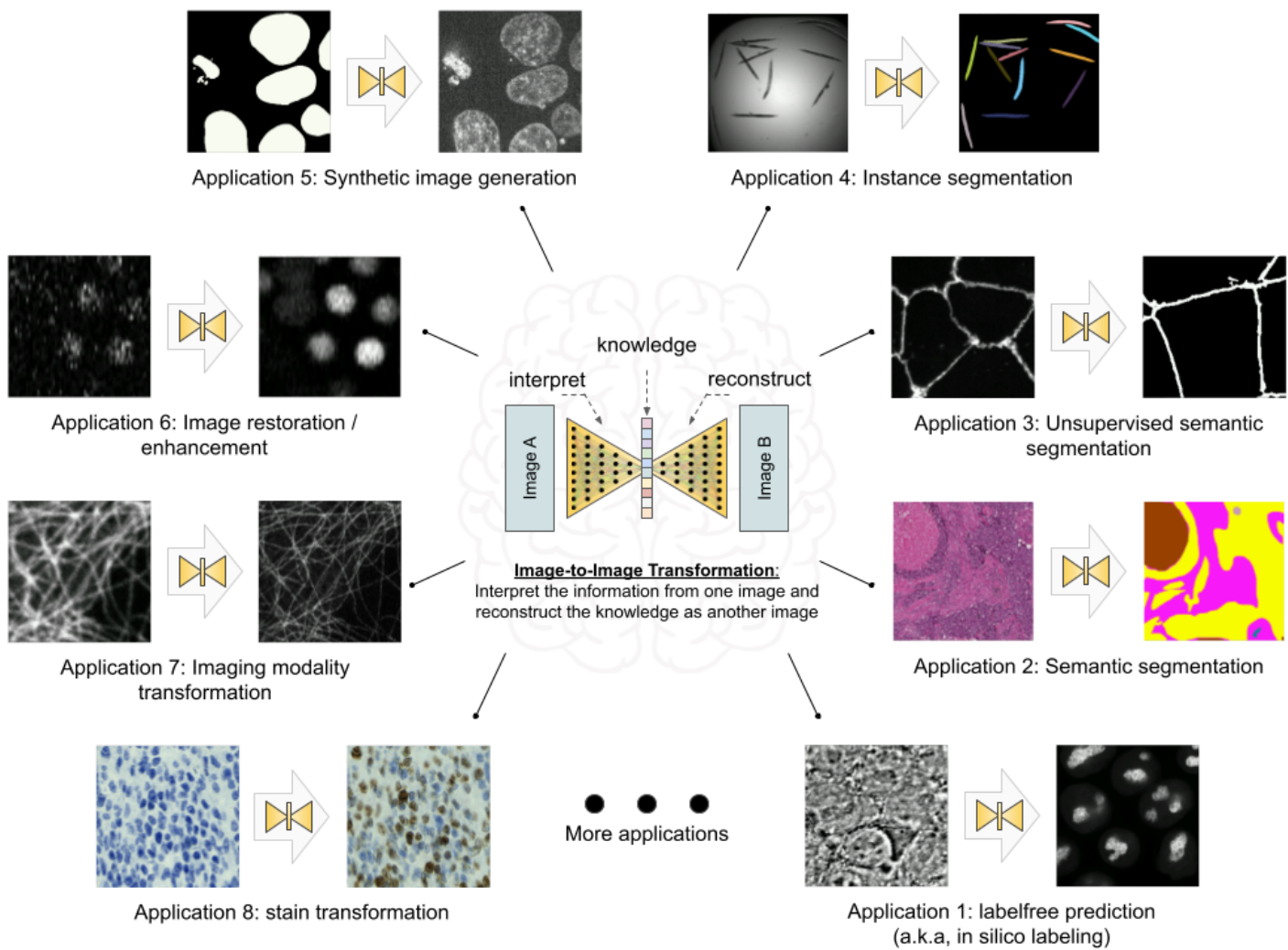

**Figure 1:** Overview of the image-to-image transformation concept and its example applications.

# Results

In this section, we showcase the versatility of the *MMV_Im2Im* toolbox by presenting over ten different biomedical applications across various R&D use cases and scales. All experiments and results in this section were conducted on publicly available datasets released with other publications and our scripts (for pulling the public dataset online and data wrangling) and configuration files (for setting up training and inference details), both included in the MMV_Im2Im package. Our aim is to make it easy to reproduce all of the results in this paper, and more importantly use these data and scripts to get familar with the package and adapt to new problems of users' interest. It is important to note that the aim of these experiments was not to achieve the best performance on each individual task, as this may require further hyper-parameter tuning (see Discussion section for more details). Rather, the experiments were intended to demonstrate the package's different features and general applicability, providing a holistic view of image-to-image transformation concepts to biomedical researchers. We hope that these concepts will help researchers integrate AI into traditional assay development strategies and inspire computational and experimental co-design methods, enabling new biomedical studies that were previously unfeasible.

## Labelfree prediction of nuclear structure from 2D/3D brightfield images

The labelfree method refers a DL method that can predict fluorescent images directly from transmitted light brightfield images [4]. Comparing to brightfield images, fluorescent images can resolve subcellular structures in living cells at high resolution but with the cost of expensive and slow procedures and high phototoxicity. The labelfree method provides a new perspective in assay development to conduct integrated computational analysis of multiple organelles only with a single brightfield image acquisition. In our first demonstration, we applied *MMV_Im2Im* to build 2D/3D models that can predict fluorescent images of nuclear structures from brightfield images. For 3D models, we also compared (1) different image normalization methods, (2) different network backbones, and (3) different types of models.

It should be noted that while we recognize the importance of systematically evaluating the predictions, such an analysis falls outside the scope of this paper. We argue that an appropriate evaluation methodology should depend on specific downstream quantitative analysis goals (e.g., [21,22]). For example, if our aim is to quantify the size of nucleoli, we must compare the segmentation derived from real nucleoli signals to that of the predicted nucleoli segmentation, ensuring that measurements from both are consistent. Alternatively, if the goal is to localize the nucleoli roughly within the cell, Pearson correlation may be a more appropriate metric. In this work, we concentrate on visual inspection, using Pearson correlation and structural similarity as a rough quantitative reference. Our intent is to demonstrate the utility of our *MMV_Im2Im* package, and leave appropriate evaluations to users in their specific problems in real studies.

*2D Labelfree:* We started with a simple problem using 2D images from the HeLa "Kyoto" cells dataset [23]. For all images, we took the brightfield channel and the mCherry-H2B channel out of the multi-channel timelapse movies. 2D images were acquired at 20x with 0.8 N.A. and then downscaled by 4 (pixel size: 0.299 nm x 0.299 nm). Example predictions can be found in Figure 2-A. We compared a basic UNet model [9] and a 2D version of the fnet model in [4]. The fnet model achieved slightly more accurate predictions than the basic UNet.

*3D Labelfree:* We tested with 3D images from the hiPS single cell image dataset [24]. Specifically, we extracted the brightfield channel and the structure channel from the full field-of-view (FOV) multi-channel images, from the HIST1H2BJ, FBL, NPM1, LMNB1 cell lines, so as to predict from one brightfield image various nuclear structures, histones, nucleoli (dense fibrillar component via fibrillarin), nucleoli (granular component via nucleophosmin), and nuclear envelope, respectively. Images were acquired at 100x with 1.25 NA (voxel size: 0.108 micron x 0.108 micron x 0.29 micron).

We conducted three groups of comparisons (see results in Figure 2-B). First, we compared three different image normalization methods for 3D images: percentile normalization, standard normalization, center normalization [4]. Percentile normalization refers to cutting the intensity out of the range of [0.5, 99.5] percentile of the image intensity and then rescale the values to the range of [-1, 1], while the standard normalization is simply subtracting mean intensity and then divided by the standard deviation of all pixel intensities. Center normalization is similar to standard normalization, but the statistics are calculated only around center along the Z-axis [4]. One could easily test different percentile or rescaling to [0, 1] instead of [-1, 1]. Qualitatively, we found center normalization slightly more accurate and more robust than the other two (ref. the first row in Figure 2-B).

Second, we compared different network backbone architectures, including the original fnet model [4], an enhanced UNet [25], the attention UNet [26], two transformer-based models, SwinUNETR [27] and UNETR[28] (all with center normalization). Inspecting the predictions on a holdout validation set suggested that fnet achieved the best performance, and the transformer-based models did not work well in labelfree problems (ref. the second row and the "c + fnet" from the first row in Figure 2-B).

Finally, we showed the comparison between three different types of models, an FCN-type model (i.e., fnet), a pix2pix-type model, and a cycleGAN-type model. For fair comparison, we used fnet as the same backbone for all three types of models. In theory, the pix2pix-type model can be trained in two different ways: from scratch or initializing the generator with a pre-trained fnet (trained as FCN).

Examples of the comparison results were shown in the last two rows in Figure 2-B. Visually, it is evident that the additional adversarial components (i.e., the discriminator) could generate images with more realistic apprearnce than a typical FCN-type model alone, but again, we leave the appropriate quantitative evaluations to users' specfic biomedical studies.

From the experiments above, we found that center normalization + pix2pix with fnet as the generator achieved the best overall performance qualitatively. So, we employed the same strategy on all other nuclear structures. At the end, we had four different labelfree models, each predicting one different nuclear structure from 3D brightfield images. As an example of evaluation, we calculated the pearson correlation and structural similarity on hold-out validation sets. The results were summarized in Table 1. Again, these numbers were merely examples of evaluation, systematic evaluation based on each specific biological problem would be necessary before deployment. Figure 2-C showed one example of all four different structures predicted from a single unseen brightfield image. This would permit an integrated analysis of four different nuclear components that could hardly be acquired simultaneously in real experiments and real images.

**Table 1:** Evaluation of the final 3D label-free models for four different nuclear structures.

| Dataset | Pearson Correlation | Structural Similarity | # of Test Data |
| --- | --- | --- | --- |
| FBL | 0.864 ± 0.021 | 0.831 ± 0.034 | 50 |
| HIST1H2BJ | 0.825 ± 0.034 | 0.675 ± 0.073 | 55 |
| LMNB1 | 0.853 ± 0.027 | 0.669 ± 0.059 | 50 |
| NPM1 | 0.912 ± 0.015 | 0.795 ± 0.039 | 55 |

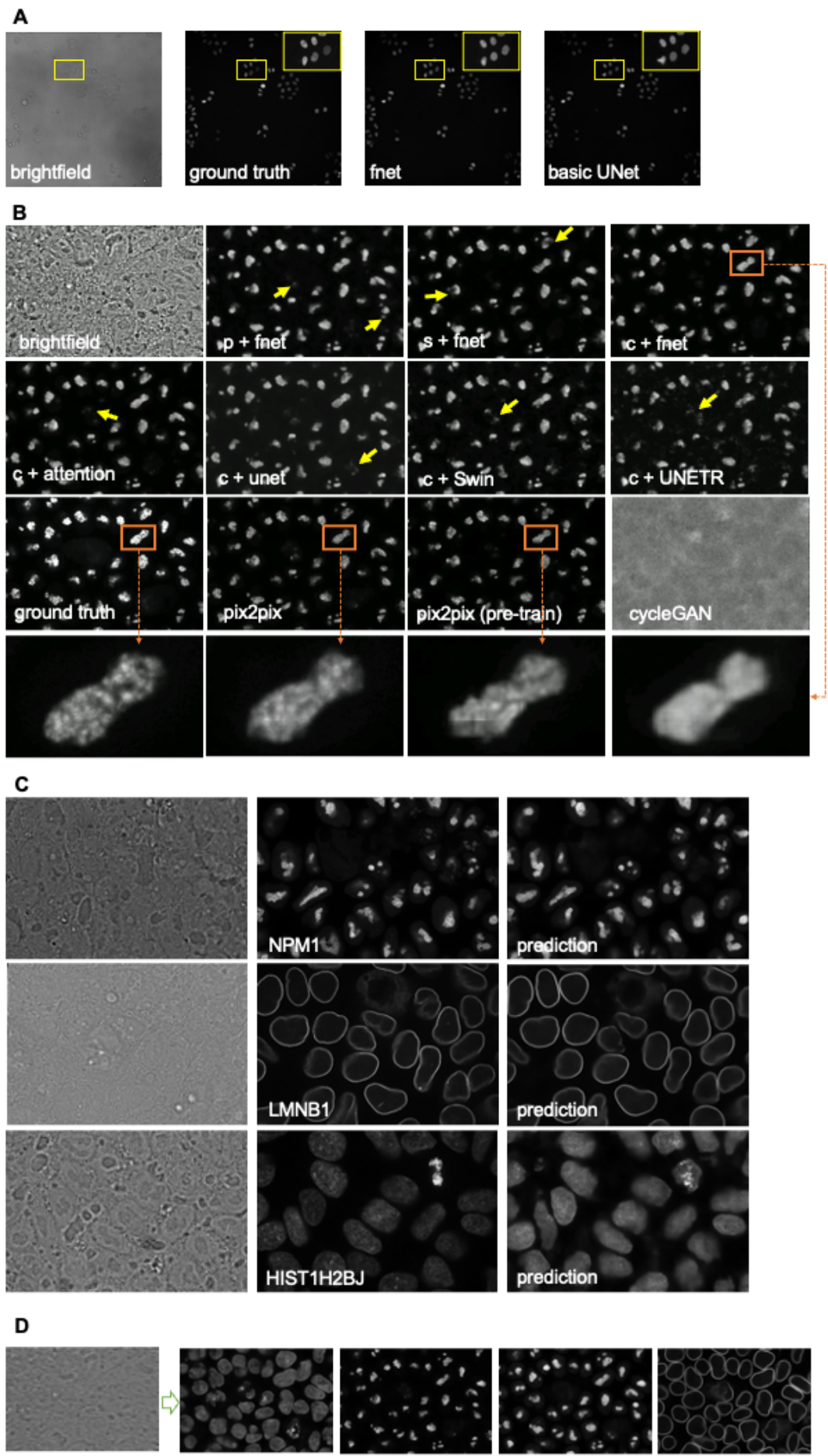

**Figure 2:** Examples of labelfree results. p/c/s refers percentile normalization, center normalization, and standard normalization, respectively (see main text for details).

## 2D semantic segmentation of tissues from H&E images

Segmentation is a common image processing task, and can be considered as a special type of image-to-image transformation, where the generated images are segmentation masks. DL based methods have achieved huge success in 2D semantic segmentation in biomedical images. In this example, we demonstrated *MMV_Im2Im* on a pathology application to segment glands from hematoxylin and eosin (H&E) stained tissue images from the 2015 Gland Segmentation challenge [29,30]. Stain normalization is an important pre-processing step in order to develop models robust to stain variation and tissue variations. *MMV_Im2Im* included a classic stain normalization method [31] as a pre-processing step. The effect of stain normalization can be observed in Figure 3-A and B. We trained a simple attention UNet model [26]. Evaluated on the two different hold-out test sets, the model achieved F1-score, 0.883 and 0.888 on test set A and test set B, respectively. The performance was competitive comparing to the methods reported in the challenge report [30], especially with much more consistent performance across the two different test sets. Example results can be found in Figure 3-C.

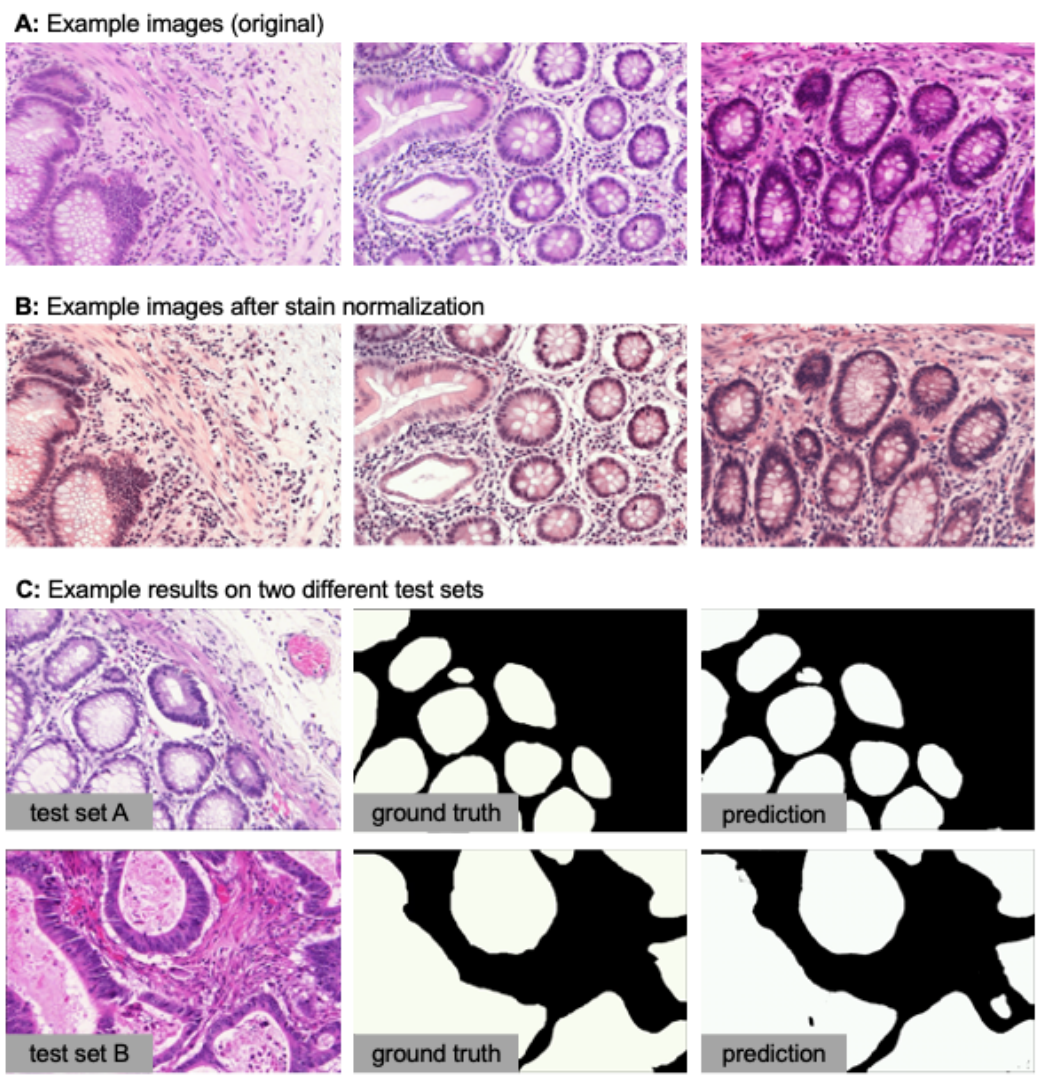

**Figure 3:** Example results of 2D semantic segmentation of gland in H&E images.

## Instance segmentation in microscopy images

Instance segmentation is a type of segmentation problem that goes beyond semantic segmentation. The goal is to differentiate not only between different types of objects, but also different instances of the same type of objects. Currently, the *MMV_Im2Im* package supports *EmbedSeg*-type models. The major benefit of EmbedSeg-type models is their agnosticism to the morphology and dimensionality of the object instances, compared to other models such as StarDist [32,33], SplineDist [34] and CellPose

[35]. For example, different from the others, *EmbedSeg*-type models are even able to generate instance segmentation where each instance contains multiple connected components. Additional frameworks such as OmniPose [36] will be supported in future versions. Another mainstream category of instance segmentation methods are detection-based models, such as Mask-RCNN [37]. However, these models are better suited to the detection framework rather than image-to-image transformation (see Discussion section for details).

The *EmbedSeg*-type models were re-implemented according to the original paper [10,11] following the generic boilerplate in *MMV_Im2Im*, with significant improvment. First of all, following the modular design of *MMV_Im2Im*, it is flexible to use different neural network models as the backbone. For 3D anisotropic microscopy image, the original backbone ERFNet [38] doesn't take the anisotropic dimensions into account and therefore may not perform well or even not applicable. In this scenario, it is straighforward to employ another anisotropic neural network bone, such as the anisotropic U-Net in [19] or the anisotropic version of Dynamic U-Net in MONAI. Second, we significantly improve training strategy. The original version requires pre-cropping patches centered around each instance and pre-calculated the center images and class images. This may generate a massive amount of additional data on the disk. More importantly, such pre-cropping makes data augmentation nearly impossible, except the simple ones like flipping (otherwise, the pre-calculated centers might be wrong), and also greatly undersamples around negative cases (e.g., background). For example, we have observed that for an Embedseg model training only with patches centered around instances, the model may suffer from degraded performance during inference when there are a large amound of background areas without any instances. Again, following the modular design of *MMV_Im2Im*, it is now possible to do on-the-fly data augmentation and patch sampling, even weighted patch sampling. Third, our improved *Embedseg*-type models can accept an exclusion mask so that certain parts of the images can be ignored during training. This is especially useful for partially annotated ground truth. For large images, it could be extremely time-consuming to require every single instance to be annotated. The exclusion masks can address this bottleneck. Another extension comparing to the original implementation was that the *MMV_Im2Im* package made sliding windowing inference straightforward, and therefore permitted easy handling of images of any size during inference.

In this work, we tested on both 2D and 3D instance segmentation problems. Going from 2D to 3D is not a simple generalization from 2D models by switching 2D operations with 3D operations, but with many practical challenges. Large GPU footprint is one of the biggest issues, which makes many training strategies common in 2D not feasible in 3D, e.g. limited mini-batch size. *MMV_Im2Im* is able to take advantage of state-of-the-art ML engineering methods to efficiently handle 3D problems. For example, by using effective half-precision training, one can greatly reduce GPU memory workload for each sample and therefore increase the batch size or the patch size. When multiple GPUs are available, it is also possible to easily take advantage of the additional resources to scale up the training to multiple GPU cards, even multiple GPU nodes. As a demonstration, we applied *EmbedSeg*-like models to a 2D problem of segmenting *C. elegans* from widefield images [39], as well as a 3D problem of nuclear segmentation from fluorescent and brightfield images from the hiPS single-cell image dataset [24].

For the 2D problem, we adopted the same network backbone as in the original *EmbedSeg* paper. Example results on a small holdout set are shown in Figure 4-A (IoU = 0.86), which is comparable to the original published results [11]. For the 3D problem, the original backbone is not directly applicable, due to the aforementioned anisotropic issue and the images in the dataset do not contain enough Z-slices to run through all down sampling blocks in 3D. The anisotropic UNet [19] is used here. The segmentation results obtained from the public dataset [24] contain nuclear instance segmentation of all cells. But, the cells touch the image borders are ignored from downstream analysis [24] and therefore not curated. In other words, the segmentation from this public dataset can only be used as high-quality nuclear instance segmentation ground truth after excluding the areas covered by cells touching the image borders [24]. Therefore, the exclusion masking function in *MMV_Im2Im* is very helpful in this example.

Example results were presented in [4](#)-B. The green box highlighted a mitotic cell (the DNA signals forming "spaghetti" shapes). Besides roughly separating the DNA signals from background, the model was also able to correctly identify the instance identity, which would be theoretically infeasible for other instance segmentation models like StarDist or CellPose. Nuclear instance segmentation from brightfield images was much more challenging than from fluorescent images. Arguably, this could be thought of as one single model doing two transformations: predicting the DNA signals from brightfield and running instance segmentation on predicted DNA signals. From Figure [4](#)-B, it was shown that the segmentation from brightfield images was comparable to the segmentation from fluorescent images, but with two caveats. First, the performance on mitotic cells was worse than the model using fluorescent images. We hypothesized this could be due to the limited information in brightfield images for mitotic cells, compounded with limited number of mitotic cells in the whole training set (less than 10%). Second, the performance along Z dimension was also worse than the fluorescent model, as explained by the side view in Figure [4](#)-B. This could be explained by the different properties of brightfield imaging and fluorescent imaging, but would need further comprehensive studies to investigate and confirm.

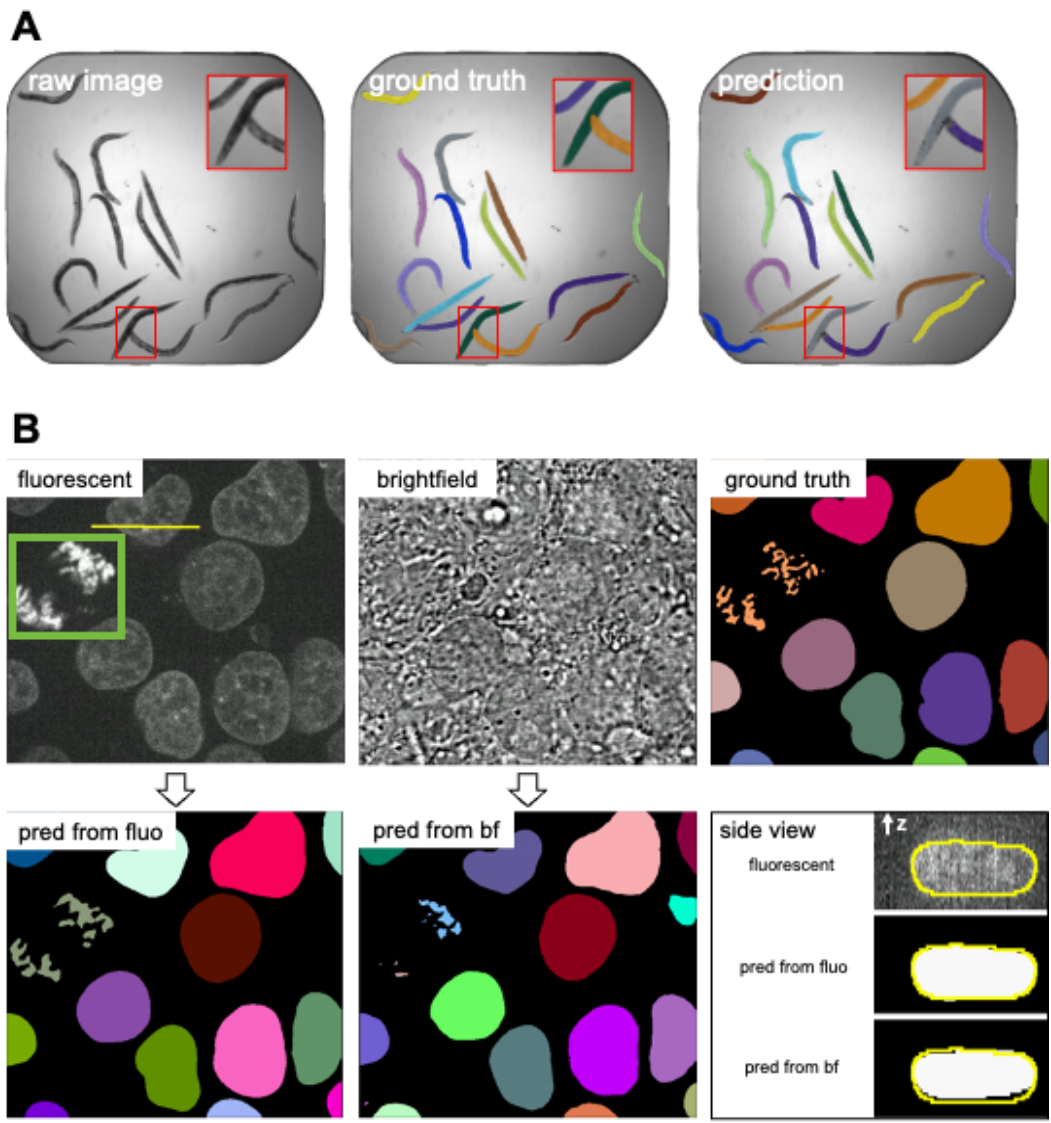

**Figure 4:** (A) Results 2D instance segmentation of C. elegans. Some minor errors can be observed in the zoom-in window, which could be refined with post-processing. (B) Results of 3D nuclear instance segmentation from fluorescent images and brightfield images. The green box in the fluorescent image highlights a mitotic example. The side view panel shows the segmentation of one specific nucleus along the line annotated in the fluorescent image from the side.

# Comparing semantic segmentation and instance segmentation of organelles from 3D confocal microscopy images

We did a special comparison in this subsection to further illustrate the difference between semantic and instance segmentations. We took the 3D fibrillarin dataset from [24]. There are multiple channels in each 3D images, including DNA dye, membrane dye, and the structure channel (i.e., fibrillarin in this case). The original fibrillarin segmentation released with the dataset is a semantic segmentation (0=background, 1=fibrillarin). With the additional cell segmentation available in the dataset, we can know which groups of segmented fibrillarin belong to the same cell. Then, we can convert the original 3D fibrillarin segmentation segmentation ground truth into 3D instance segmentation ground truth (fibrillarin pixels belonging to the same cell are grouped as a unique instance). Sample images and results are shown in Figure 5. We can observe that the semantic segmentation model is able to achieve good accuracy in determining pixels from the fibrillarin signals. Meanwhile, the instance segmentation can group them properly so that fibrillarin masks from the same cell are successfully identified as unique instances, even without referring to the cell membrane channel or cell segmentation results. This is not a simple grouping step based on distance, since the fibrillarin signals from tightly touching nuclei may exist very close to each other.

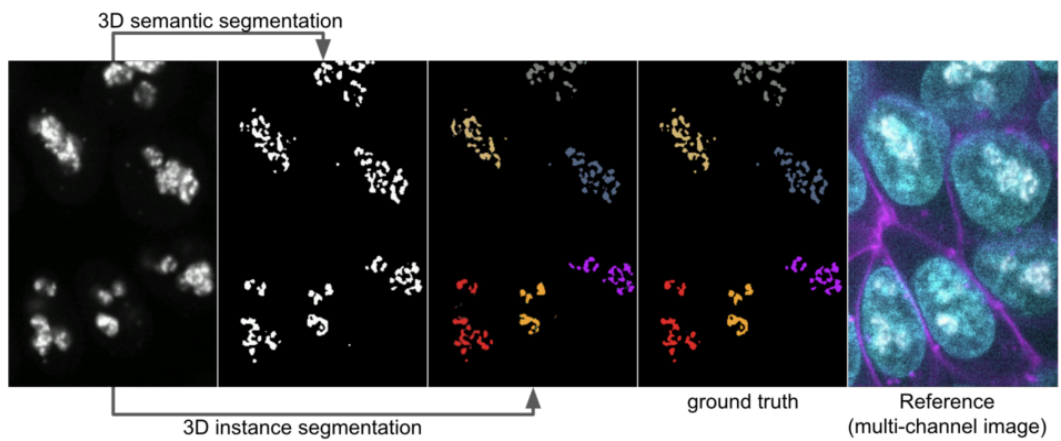

**Figure 5:** Comparing 3D semantic segmentation and 3D instance segmentation results on confocal microscopy images of fibrillarin (showing a middle Z-slice of a 3D stack). From left to right: raw fibrillarin image, semantic segmentation, instance segmentation, instance ground truth, a reference images with all the channels (DNA dye in cyan, membrane dye in magenta, and fibrillarin in white) to show all the neighering cells.

# Unsupervised semantic segmentation of intracelluar structures from 2D/3D confocal microscopy images

Large amounts of high-quality segmentation ground truth is not always available, or may require endless effort to collect for a segmentation task. CycleGAN-based methods have opened up a new avenue for segmentation without the need for pixel-wise ground truth [12]. In this subsection, we demonstrate an unsupervised learning-based segmentation method on four examples: 2D tight-junction (via ZO1) segmentation from 2D FP-tagged ZO1 images (max-projected from 3D stacks), and segmentation of nuclei, mitochondria, and golgi from 3D confocal microscopy images.

To perform unsupervised learning, we used raw images from the hiPS single-cell image dataset [24], as well as their corresponding segmentations (may not be absolute pixel-wise ground truth, but have gone through systematic evaluation to ensure the overall quality). We shuffled the raw images and their segmentations to generate a set of simulated segmentation masks. A demonstration of the concept is illustrated in Figure 6-A. Example results for all 3D models are shown in Figure 6-B, and the F1-scores on the test set are summarized in Table 2.

Interestingly, on the 2D ZO1 example, we observed that the segmentation generated by the unsupervised learning method was actually slightly better than the original segmentation obtained from a classic image segmentation workflow as in [19]. For the 3D examples, it has been suggested that the quality of unsupervised nuclei segmentation could be further improved with additional simulation strategies [12]. Overall, we believe that unsupervised learning offers an effective way to generate preliminary segmentation, which can be further refined through active learning such as the iterative DL workflow described in [19]

**Table 2:** F1 scores of the unsupervised semantic segmentation predictions.

| Dimensionality | Dataset | F1 Score | # of Test Data |
| --- | --- | --- | --- |
| 2D | tight-junction | 0.888 ± 0.022 | 29 |
| 3D | nucleus | 0.811 ± 0.150 | 15 |
| 3D | golgi | 0.705 ± 0.022 | 6 |
| 3D | mitochondria | 0.783 ± 0.005 | 2 |

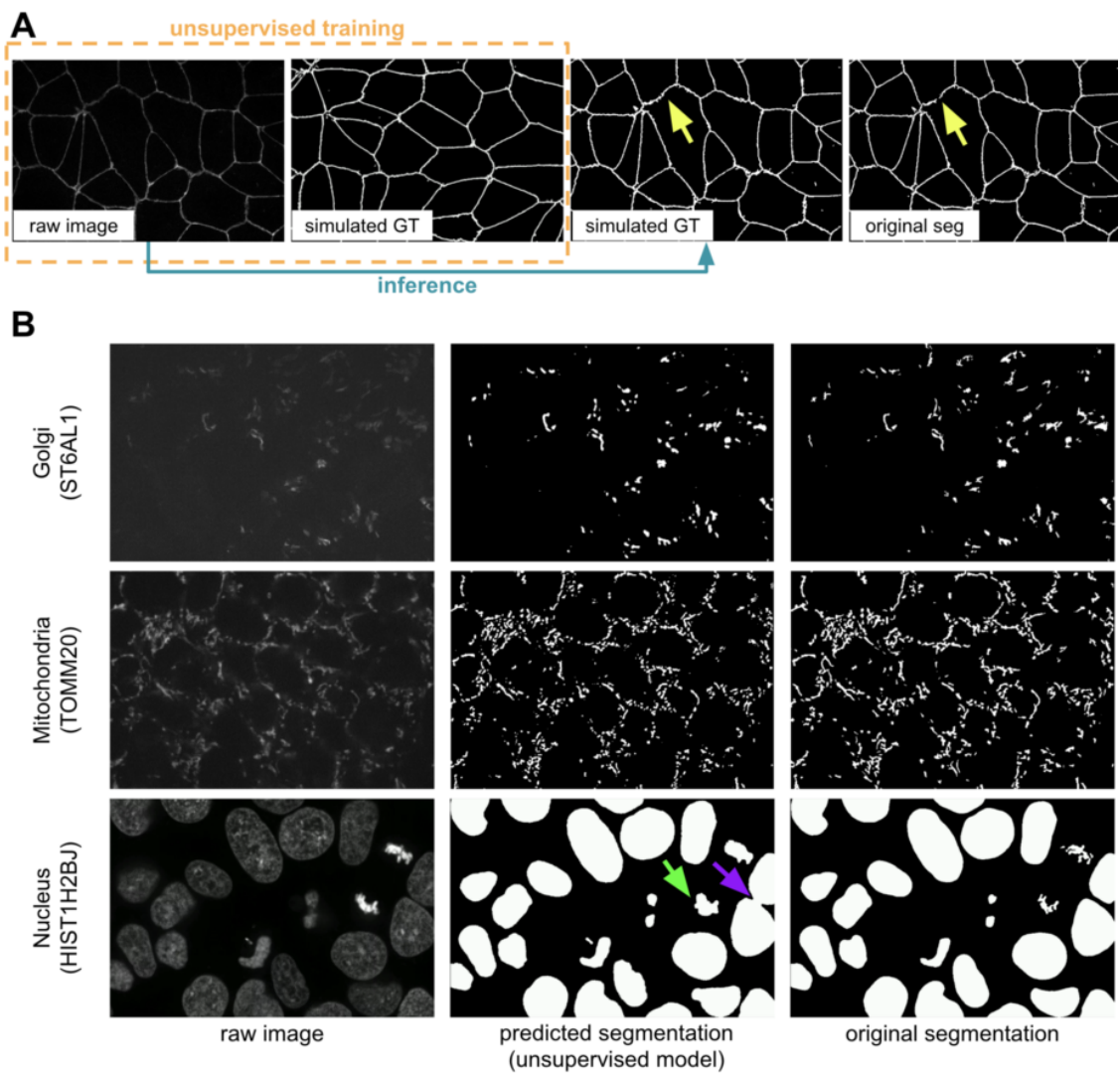

**Figure 6:** (A) Illustration of the unsupervised learning scheme and results in the 2D tight-junction segmentation problem. The yellow arrows indicate a region where the unsupervised learning method actually predicted better than the original segmentation results by classic image processing methods. (B) Example 3D segmentation results (only showing a middle z-slice) from models obtained by unsupervised learning.

# Generating synthetic microscopy images from binary Masks

Generating a large amount of synthetic microscopy images can be an important step in developing image analysis methods. Synthetic images offer a way to train other DL models, such as self-supervised pre-training, using a diverse set of images without the need for large amounts of real-world data. As long as the synthetic images are generated with sufficient quality, it is possible to have an unlimited amount of training data for certain applications. Moreover, synthetic images can be used to evaluate other models when validation data is difficult to obtain. In this study, we demonstrate that *MMV_Im2Im* can generate 2D/3D synthetic microscopy images with high realism and validity, using a subset of data collected from the hiPS single-cell image dataset [24], either in a supervised or unsupervised manner.

For 2D demonstration, we extracted the middle Z-slice from NPM1 images as the training target, while using the NPM1 segmentation results as the input binary mask. With the paired "mask + microscopy image" data, we could train the model in a supervised fashion, or randomly shuffle the data to simulate the situation without paired data which can be trained in an unsupervised fashion. Example results can be found in Figure 7-A. In general, the supervised synthesization can generate more realistic images than the unsupervised model.

For 3D demonstration, we use 3D H2B images with two different types of input masks. First, we attempted to generate synthetic images from a coarse mask (i.e., only the overall shape of the nucleus, available as nuclear segmentation from the dataset) with both supervised training and unsupervised training. The unsupervised model in *MMV_Im2Im* uses the CycleGAN-based approaches. So, the unsupervised training is actually already done within the unsupervised segmentation experiments. In other words, the unsupervised model works in a bi-directional way, from real microscopy images to binary mask, and also from binary masks to simulated microscopy images. Here, we just do the inference in a different direction (from binary to simulated microscopy) using the model trained in unsupervised segmentation section. The results are shown in Figure 7-B (row 1). Similar to the results in 2D demonstration, the unsupervised synthesization can mostly "paint" the mask with homogeneous grayscale intensity, while the supervised model can simulate the textures to some extent. For a relatively large mask, it could be chanllenging for a model to fill in sufficient details to simulate real microscopy images (might be improved with diffusion-based models, see Discussions).

We made another attempt with 3D masks containing finer details beyond the overall shapes. So, we employed the H2B structure segmentation results from the dataset (capturing the detailed nuclear components marked by histon H2B) as the input for supervsied synthesization. The result is shown in Figure 7-B (row 2). Comparing to the synthesization with coarse masks, the images simulated from fine masks exhibit much more realistic appearance. As we can see, it is important to design the solutions with proper data.

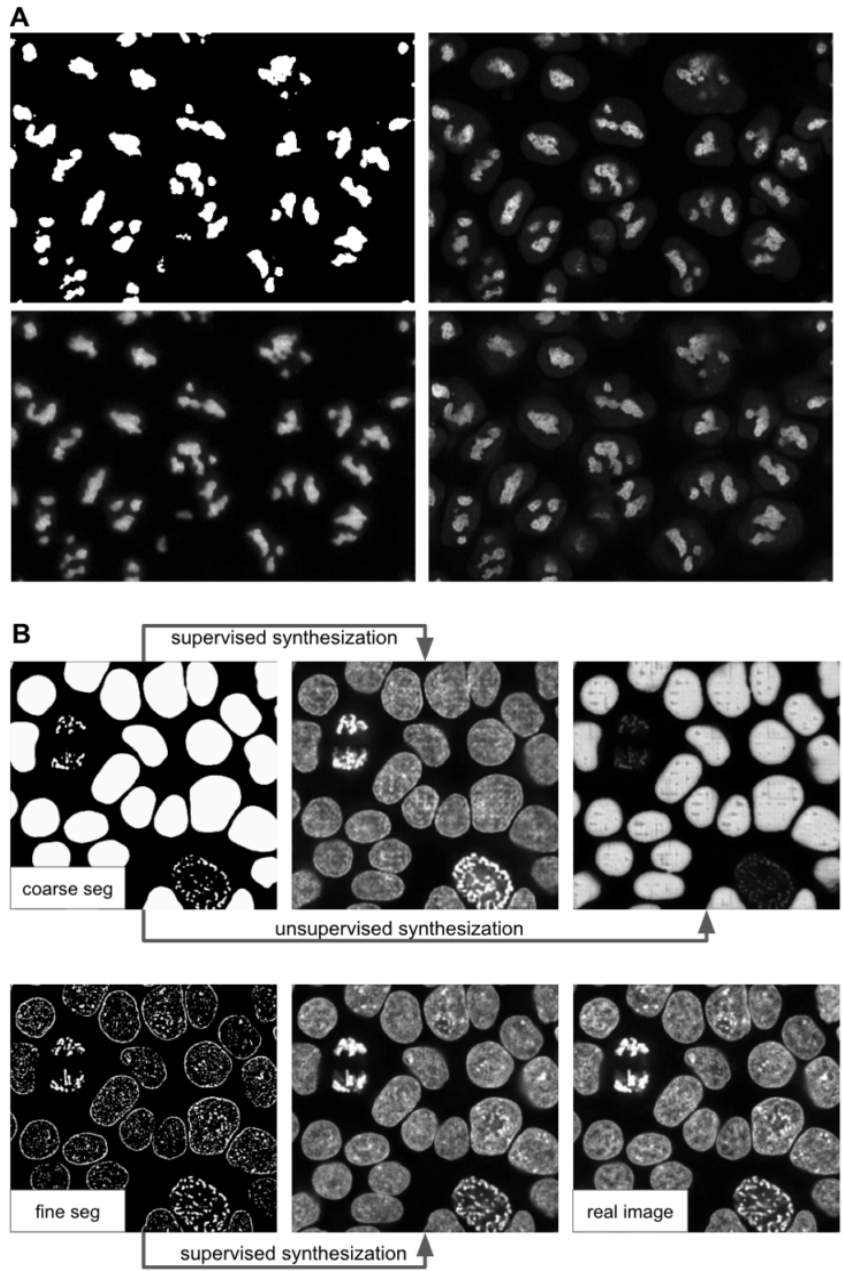

**Figure 7:** Example results of (A) 2D synthetic fluorescent images of nucleoli (via NPM1) and (B) 3D synthetic fluorescent images of H2B (middle z-slices of a z-stack) with a coarse mask and a fine mask as the input.

## Image denoising for microscopy images

*MMV_Im2Im* can also be used to computationally reduce image noise or restore the data from various sources of imaging artifacts, so as to increase the feasibility and efficiency in downstream analysis. In the current version of *MMV_Im2Im*, the restoration model can only be trained in a fully supervised manner. Therefore, aligned low quality and high quality images are required for supervision, even though such pair data can be partially simulated [6]. Other methods, such as unsupervised learning based solutions [41], will be made available within *MMV_Im2Im* in future versions.

In this example, we presented an image denoising demonstration with sample data from [42]. The goal was to increase the quality of low signal-to-noise ratio (SNR) images of nucleus-stained flatworm (Schmidtea mediterranea) and lightsheet images of Tribolium castaneum (red flour beetle) embryos. The models were trained with paired data acquired with low and high laser intensity on fixed samples, and then applied on live imaging data. For the nucleus-stained flatworm data (a test set of 20 images are available), the model achieved pearson correlation of 0.923 ± 0.029 and structural similarity of 0.627 ± 0.175. Based on the results in Figure 8, it can be observed that the low SNR images can be greatly improved. Systematic quantitative evaluations would be necessary to confirm the biological validity, but beyond the scope of this paper.

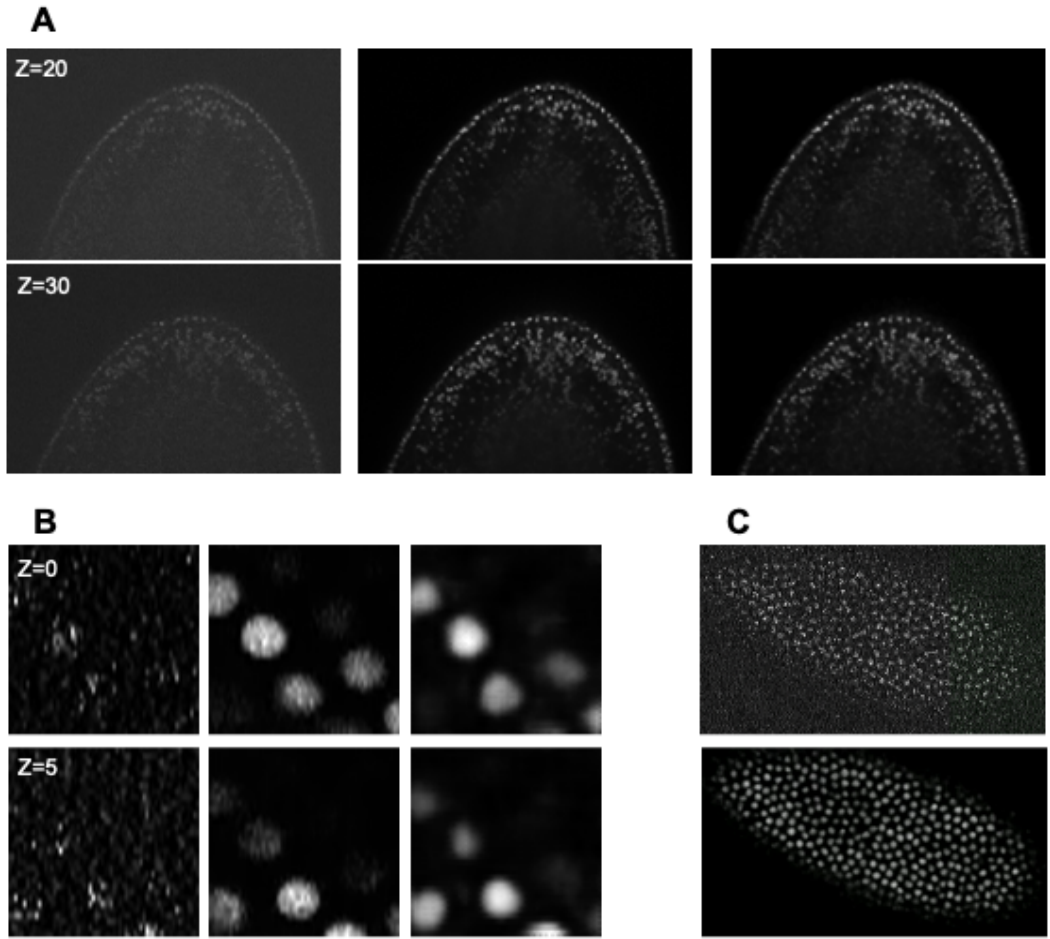

**Figure 8:** (A) Denoising results of 3D images of nucleus-stained flatworm at two different z-slices. Left: raw images (low SNR), middle: reference images (high SNR), right: predictions. (B) Denoising results of 3D lightsheet images of Tribolium castaneum (fixed samples) at two different z-slices. Left: raw images (low SNR), middle: Reference images (high SNR), right: predictions. (C) Denoising results of 3D lightsheet images of Tribolium castaneum (live samples) without high SNR reference. Top: the raw image, bottom: the prediction.

# Imaging modality transformation from 3D confocal microscopy images to stimulated emission depletion (STED) microscopy images

Another important application of image-to-image transformation is imaging modality transformation [8], usually from one "cheaper" modality with lower resolution (e.g., with larger field-of-view, easier to acquire and scale up) to another modality with higher resolution but expensive to obtain. Such models will permit a new way in assay development strategy to take advantage of all the benefits of the cheaper modality with lower resolution and still able to enhance the resolution computationally post hoc. To demonstrate the application of *MMV_Im2Im* in this scenario, we took an example dataset with paired 3D confocal and Stimulated Emission Depletion (STED) images of two different cellular structures, microtubule and nuclear pore [8]. Sample results were summarized in Figure 9. For microtubule, the model achieved pearson correlation of 0.779 ± 0.019, while for nuclear pore complex, the pearson correlation was 0.784 ± 0.028. Also, visual inspection can confirm the effectiveness of the models. Again, it would be necessary to conduct further quantitative evaluation to ensure the validity in users' specific problems.

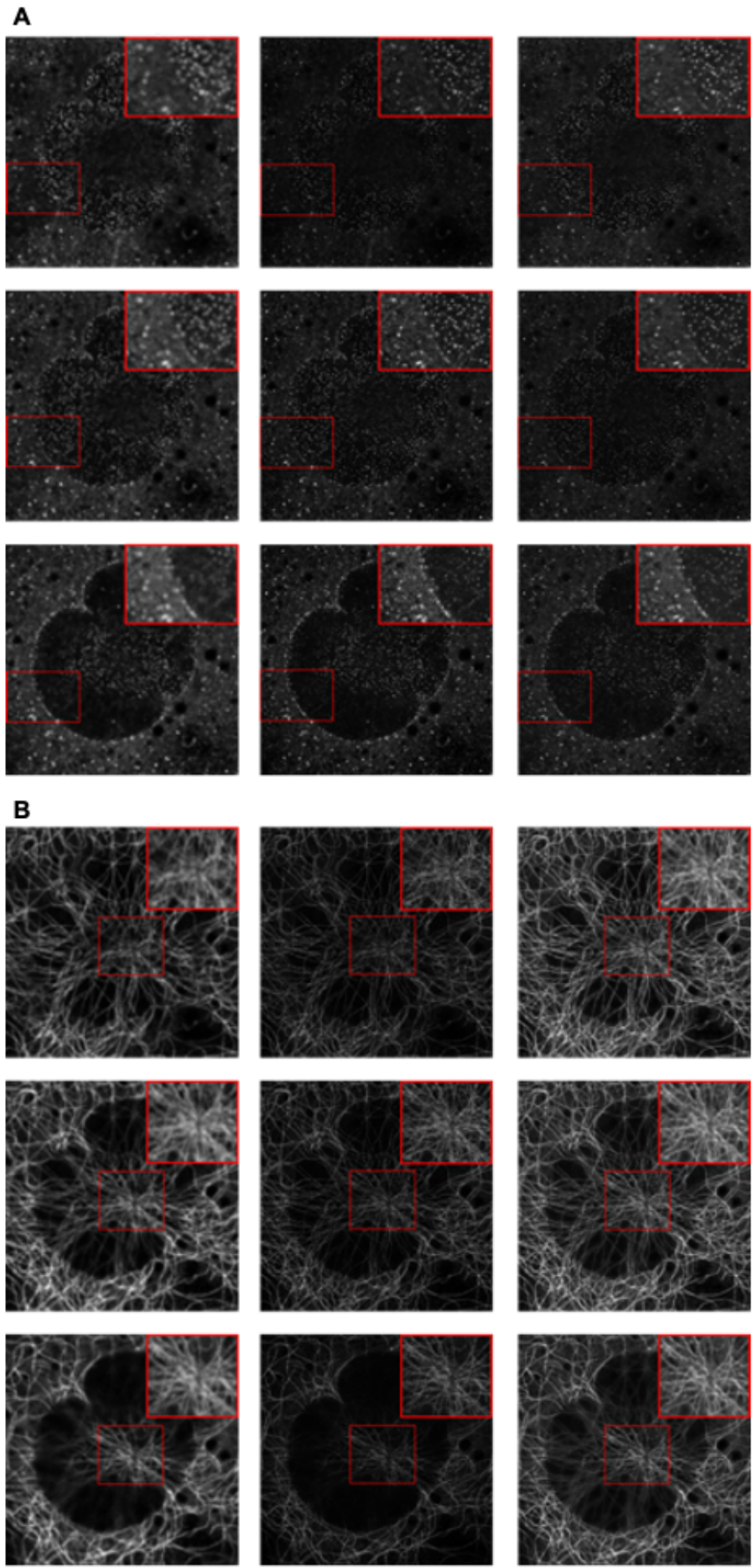

**Figure 9:** Example results of confocal-to-STED modality transformation of nuclear pore (A) and microtubule (B) in three consecutive z-slices. From left to right: raw confocal images, reference STED images, predicted images.

## Staining transformation in multiplex experiments

DL has emerged as a powerful tool for multiplex imaging, a powerful technique that enables the simultaneous detection and visualization of multiple biomolecules within a single tissue sample. This technique is increasingly being used in biomedical expereiements but demands efficient image analysis solutions to accurately identify and quantify the different biomolecules of interest at scale. DL has demonstrated great potentials in analyzing muliplex datasets, as it can automatically learn the complex relationships between different biomolecules and their spatial distribution within tissues. Specially, in this study, we present the effectiveness of *MMV_Im2Im* in transforming tissue images from one staining to another, which will permit efficient co-registration, co-localization, and quantitative analysis of multiplex datasets. We used the sample dataset from [5]. In this example, we trained three different models to transform IHC images to images of standard hematoxylin stain, mpIF nuclear (DAPI) and mpIF LAP2beta (a nuclear envelope stain). Example results can be observed in Figure 10 to verify the results qualitatively. These transformed images can provide valuable insights into the localization and expression patterns of specific biomolecules spatially.

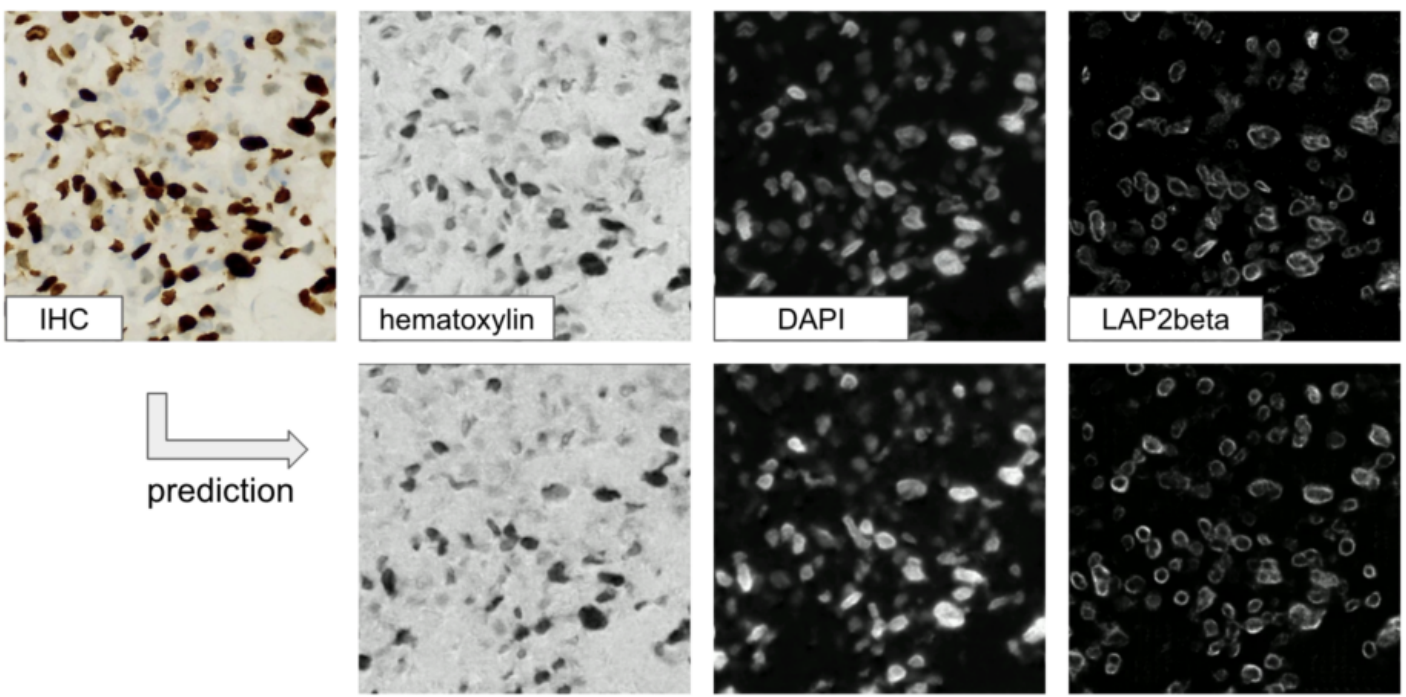

**Figure 10:** Qualitative visualization of staining transformation results with the *MMV_Im2Im* package.

# Methods

## Overview of the code base

Overall, the package inherited the boilerplate concept from pytorch-lightning (https://www.pytorchlightning.ai/), and was made fully configurable via yaml files supported by pyrallis (https://github.com/eladrich/pyrallis), as well as largely employed state-of-the-art DL components from MONAI (https://monai.io/). The three key parts in the package: mmv_im2im.models, mmv_im2im.data_modules, and Trainers, will be further described below.

## Main frameworks for mmv_im2im.models

*mmv_im2im.models* is the core module defining the DL framework for your problem, where we can instantiate the neural network architecture and define what to do before training starts, what to do in each training and validation step, what to do at the end of each epoch, etc.. All implemented following the same lightning module from pytorch-lightning, which makes the code very easy to read, to understand, and even to extend.

In general, there are mainly four major DL frameworks that could be applied to microscopy image-to-image transformation: supervised learning with a fully convolutional networks (FCN) type models, supervised learning with pix2pix type models, unsupervised learning to learn mapping between visual domains, and Self2Self-type self-supervised learning [43]. The major difference between FCN based supervised learning and pix2pix based supervised learning is that the pix2pix framework extends an FCN model with an adversarial head as a discriminator to further improves the realism of the prediction. The major difference between the unsupervised framework and the self-supervised framework is that the unsupervised methods still requires examples of the target images, even though the source images and target images do not need to be from the same sample or pixel-wise aligned. But, the self-supervised framework would only need the original images, which could be really helpful when it is impossible to acquire the target images (e.g., there is no truely noise-free or artifact-free image).

Currently, for supervised frameworks, both the FCN-type and pix2pix-type are well supported in the *MMV_Im2Im* package. Since our package is designed in a very generic way, it is possible to continuously expand the functionalities when available (ideally with community contributions). For example, diffusion models [44] can be thought of a modern extension of the pix2pix-type framework and therefore are within our horizon to include into *MMV_Im2Im*. For the unsupervised framework, only CycleGAN-type methods are supported. We are planning to extend the unsupervised framework with Imaginaire (https://github.com/NVlabs/imaginaire), which will greatly extend the applicability of *MMV_Im2Im* (e.g., learning the transformation from one single image to another single image or one set of images to another set of images). Meanwhile, supporting the self-supervised framework will be our next major milestone.

## Customized mmv_im2im.data_modules for bioimaging applications

The *data_modules* implements a general module for data handling for all differnet frameworks mentioned above, from how to load the data to how to set up the dataloader for training and validation. Different people may prefer to organize their training data in different ways, such as using csv to organize input and the corresponding ground truth, or making different folders (e.g. "image" and "ground_truth") with input and the corresponding ground truth sharing the same file name, etc.. Or some people may prefer to do a random train/validation split, while others like to pre-split train and validation into different folders, etc.. Currently, the *data_module* in *MMV_Im2Im* supports four different ways of data loading, where we try to cover as many common scenario as possible, so that everyone will feel comfortable using it.

A big challenge in the dataloader in bioimaging applications is that there could be not only a large amount of files, but also files of very large sizes. To deal with each individual large image, we used the delayed loading from aicsimageio for efficient image reading. Besides, we adopted the PersistentDataloader from MONAI to further optimize the efficiency. In specific, after loading a large image and running through all the deterministic operations, like intensity normalization or spatial padding, the PersistentDataLoader will pickle and save the data in a temporary folder, to avoid repeating the heavy computation on large files in each training iteration. To handle the potentially large number of files, we implemented the data_module with capability of loading only a certain portion of the data into the memory in each epoch and reloading with a different portion every certain number of epochs. By doing this, we were able to efficiently train an instance segmentation model with more than 125K images, where each raw image is about 15MB.

## State-of-the-art training with the pytorch-lightning Trainer

We fully adopted the Trainer from pytorch-lightning, which has been widely used by the machine learning community, and wildly tested on both R&D problems and industrial-scale applications. In a nutshell, simply by specifying the training parameters in the yaml file, users can setup multi-GPU training, half-precision training, automatic learning rate finder, automatic batch size finder, early stopping, stochastic weight averaging, etc.. This allows users to focus on the research problems without worrying about the ML engineering.

## Discussions

In this work, we presented a new open source python package *MMV_Im2Im* package for image-to-image transformations in bioimaging applications. We demonstrated the applicability on more than ten different problems or datasets to give biomedical researchers a holistic view of the general image-to-image transformation concepts with diverse examples. This package was not a simple collection of existing methods. Instead, we distilled the knowledge from existing methods and created this generic version with state-of-the-art ML engineering techniques, which made the package easy to understand, easy to use, and easy to extend for future. We hope this package can serve the starting point for other researchers doing AI-based image-to-image transformation research, and eventually build a large shared community in the field of image-to-image transformation for bioimaging.

One of main directions for extending *MMV_Im2Im* is to pre-pack common bioimaging datasets as a *Dataset* module, so that DL researchers can use for algorithm development and benchmarking, and new users can easily use for learning microscopy image-to-image transformation. We will continue improving the functionalities of the package, such as supporting more models and methods, such as diffusion based models [44], unsupervised denoising [41] or Imaginaire (https://github.com/NVlabs/imaginaire). Besides, we also plan to develop two auxillary packages *MMV_Im2Im_Auto* and *MMV_Im2Im_Active*. In specific, when you have a reasonable amount of training data, MMV_Im2Im_Auto will take advantage of the fact that *MMV_Im2Im* is fully configurable with yaml files, and automatically generate a set of potentially good configurations, then find the optimal solution for you by cross validation. On the other hand, when you only have very limited training data, or even with only pseudo ground truth, *MMV_Im2Im_Active* will help to build preliminary models from the limited training data, and gradually refine the model with human-in-the-loop by active learning [19]. All the packages will also be wrapped into Napari plugins [45] to allow no-code operation and therefore be more friendly to users without experience in programming.

Finally, beyond *MMV_Im2Im*, we hope to build develop similar package for other problems (without re-inventing wheels). For example, as we mentioned in the instance segmentation application, Mask-RCNN type models are also very powerful instance segmentation methods and, in theory, can also be generalized beyond 2D images. However, Mask-RCNN would fit more to a detection framework, instead of image-to-image transformation. It will be supported in our *MMV_NDet* (NDet = N-dimensional detection) package, currently under development.

## Code and data availability

Project name: MMV_Im2Im (Microscopy Machine Vision, Image-to-Image transformation)

Project home page: https://github.com/mmv-lab/mmv_im2im

Operating system(s): Linux and Windows (when using GPU), also MacOS (when only using CPU)

Programming language: Python

## Acknowledgments

We would like to thank the MONAI team for their support in our process of development, and the aicsimageio team for advices on how to integrate aicsimageio into the package. This work is supported by the Federal Ministry of Education and Research (Bundesministerium für Bildung und Forschung, BMBF) under the funding reference 161L0272.

# References

1.  **Enhanced Deep Residual Networks for Single Image Super-Resolution**
    Bee Lim, Sanghyun Son, Heewon Kim, Seungjun Nah, Kyoung Mu Lee
    *2017 IEEE Conference on Computer Vision and Pattern Recognition Workshops (CVPRW)* (2017-07) https://doi.org/gfxhp3
    DOI: 10.1109/cvprw.2017.151

2.  **Image-to-Image Translation with Conditional Adversarial Networks**
    Phillip Isola, Jun-Yan Zhu, Tinghui Zhou, Alexei A Efros
    *2017 IEEE Conference on Computer Vision and Pattern Recognition (CVPR)* (2017-07)
    https://doi.org/gfrfv9
    DOI: 10.1109/cvpr.2017.632

3.  **Panoptic Segmentation**
    Alexander Kirillov, Kaiming He, Ross Girshick, Carsten Rother, Piotr Dollar
    *2019 IEEE/CVF Conference on Computer Vision and Pattern Recognition (CVPR)* (2019-06)
    https://doi.org/ggp9z4
    DOI: 10.1109/cvpr.2019.00963

4.  **Label-free prediction of three-dimensional fluorescence images from transmitted-light microscopy**
    Chawin Ounkomol, Sharmishtaa Seshamani, Mary M Maleckar, Forrest Collman, Gregory R Johnson
    *Nature Methods* (2018-09-17) https://doi.org/gd7d5f
    DOI: 10.1038/s41592-018-0111-2 · PMID: 30224672 · PMCID: PMC6212323

5.  **Deep learning-inferred multiplex immunofluorescence for immunohistochemical image quantification**
    Parmida Ghahremani, Yanyun Li, Arie Kaufman, Rami Vanguri, Noah Greenwald, Michael Angelo, Travis J Hollmann, Saad Nadeem
    *Nature Machine Intelligence* (2022-04-07) https://doi.org/gqc7gd
    DOI: 10.1038/s42256-022-00471-x · PMID: 36118303 · PMCID: PMC9477216

6.  **Deep learning-based point-scanning super-resolution imaging**
    Linjing Fang, Fred Monroe, Sammy Weiser Novak, Lyndsey Kirk, Cara R Schiavon, Seungyoon B Yu, Tong Zhang, Melissa Wu, Kyle Kastner, Alaa Abdel Latif, … Uri Manor
    *Nature Methods* (2021-03-08) https://doi.org/gjhgrw
    DOI: 10.1038/s41592-021-01080-z · PMID: 33686300 · PMCID: PMC8035334

7.  **LIVECell—A large-scale dataset for label-free live cell segmentation**
    Christoffer Edlund, Timothy R Jackson, Nabeel Khalid, Nicola Bevan, Timothy Dale, Andreas Dengel, Sheraz Ahmed, Johan Trygg, Rickard Sjögren
    *Nature Methods* (2021-08-30) https://doi.org/gmptqs
    DOI: 10.1038/s41592-021-01249-6 · PMID: 34462594 · PMCID: PMC8440198

8.  **Three-dimensional residual channel attention networks denoise and sharpen fluorescence microscopy image volumes**
    Jiji Chen, Hideki Sasaki, Hoyin Lai, Yijun Su, Jiamin Liu, Yicong Wu, Alexander Zhovmer, Christian A Combs, Ivan Rey-Suarez, Hung-Yu Chang, … Hari Shroff
    *Nature Methods* (2021-05-31) https://doi.org/gkbctn
    DOI: 10.1038/s41592-021-01155-x · PMID: 34059829

9.  **U-Net: Convolutional Networks for Biomedical Image Segmentation**

Olaf Ronneberger, Philipp Fischer, Thomas Brox
*Lecture Notes in Computer Science* (2015) https://doi.org/gcgk7j
DOI: 10.1007/978-3-319-24574-4_28

10. **EmbedSeg: Embedding-based Instance Segmentation for Biomedical Microscopy Data**
    Manan Lalit, Pavel Tomancak, Florian Jug
    *Medical Image Analysis* (2022-10) https://doi.org/grxbwr
    DOI: 10.1016/j.media.2022.102523 · PMID: 35926335

11. **Embedding-based Instance Segmentation in Microscopy**
    Manan Lalit, Pavel Tomancak, Florian Jug
    *Proceedings of the Fourth Conference on Medical Imaging with Deep Learning*
    https://proceedings.mlr.press/v143/lalit21a.html

12. **Unsupervised data to content transformation with histogram-matching cycle-consistent generative adversarial networks**
    Stephan J Ihle, Andreas M Reichmuth, Sophie Girardin, Hana Han, Flurin Stauffer, Anne Bonnin, Marco Stampanoni, Karthik Pattisapu, János Vörös, Csaba Forró
    *Nature Machine Intelligence* (2019-09-16) https://doi.org/ggwbcv
    DOI: 10.1038/s42256-019-0096-2

13. **Unpaired Image-to-Image Translation Using Cycle-Consistent Adversarial Networks**
    Jun-Yan Zhu, Taesung Park, Phillip Isola, Alexei A Efros
    *2017 IEEE International Conference on Computer Vision (ICCV)* (2017-10) https://doi.org/gfhw33
    DOI: 10.1109/iccv.2017.244

14. **PyTorchLightning/pytorch-lightning: 0.7.6 release**
    William Falcon, Jirka Borovec, Adrian Wälchli, Nic Eggert, Justus Schock, Jeremy Jordan, Nicki Skafte, Ir1dXD, Vadim Bereznyuk, Ethan Harris, … Anton Bakhtin
    *Zenodo* (2020-05-15) https://doi.org/gqc7f9
    DOI: 10.5281/zenodo.3828935

15. **Averaging Weights Leads to Wider Optima and Better Generalization**
    Pavel Izmailov, Dmitrii Podoprikhin, Timur Garipov, Dmitry Vetrov, Andrew Gordon Wilson
    *34th Conference on Uncertainty in Artificial Intelligence 2018, UAI 2018*
    http://auai.org/uai2018/proceedings/papers/313.pdf

16. **ImageNet: A large-scale hierarchical image database**
    Jia Deng, Wei Dong, Richard Socher, Li-Jia Li, Kai Li, Li Fei-Fei
    *2009 IEEE Conference on Computer Vision and Pattern Recognition* (2009-06)
    https://doi.org/cvc7xp
    DOI: 10.1109/cvpr.2009.5206848

17. **Project MONAI**
    The MONAI Consortium
    *Zenodo* (2020-12-15) https://doi.org/gqc7gb
    DOI: 10.5281/zenodo.4323059

18. **AllenCellModeling/aicsimageio: Types, Filesystem Management, and BioformatsReader Fixes**
    Jackson Maxfield Brown, Jamie Sherman, Toloudis, Madison Swain-Bowden, Talley Lambert, Matte Bailey, Basu Chaudhuri, Gregory Johnson, Ian Hunt-Isaak, Nicholas-Schaub, … Peter Sobolewski
    *Zenodo* (2022-05-27) https://doi.org/gqc7gc
    DOI: 10.5281/zenodo.6585658

19. **The Allen Cell and Structure Segmenter: a new open source toolkit for segmenting 3D intracellular structures in fluorescence microscopy images**
    Jianxu Chen, Liya Ding, Matheus P Viana, HyeonWoo Lee, MFilip Sluezwski, Benjamin Morris, Melissa C Hendershott, Ruian Yang, Irina A Mueller, Susanne M Rafelski
    *Cold Spring Harbor Laboratory* (2018-12-08) https://doi.org/gkspnm
    DOI: 10.1101/491035

20. **Open collaborative writing with Manubot**
    Daniel S Himmelstein, Vincent Rubinetti, David R Slochower, Dongbo Hu, Venkat S Malladi, Casey S Greene, Anthony Gitter
    *PLOS Computational Biology* (2019-06-24) https://doi.org/c7np
    DOI: 10.1371/journal.pcbi.1007128 · PMID: 31233491 · PMCID: PMC6611653

21. **Practical fluorescence reconstruction microscopy for large samples and low-magnification imaging**
    Julienne LaChance, Daniel J Cohen
    *PLOS Computational Biology* (2020-12-23) https://doi.org/grxbww
    DOI: 10.1371/journal.pcbi.1008443 · PMID: 33362219 · PMCID: PMC7802935

22. **Understanding metric-related pitfalls in image analysis validation**
    Annika Reinke, Minu D Tizabi, Michael Baumgartner, Matthias Eisenmann, Doreen Heckmann-Nötzel, AEmre Kavur, Tim Rädsch, Carole H Sudre, Laura Acion, Michela Antonelli, … Lena Maier-Hein
    *arXiv* (2023) https://doi.org/grxbwx
    DOI: 10.48550/arxiv.2302.01790

23. **HeLa "Kyoto" cells under the scope**
    Romain Guiet
    *Zenodo* (2022-02-25) https://doi.org/gqdkdm
    DOI: 10.5281/zenodo.6139958

24. **Integrated intracellular organization and its variations in human iPS cells**
    Matheus P Viana, Jianxu Chen, Theo A Knijnenburg, Ritvik Vasan, Calysta Yan, Joy E Arakaki, Matte Bailey, Ben Berry, Antoine Borensztejn, Eva M Brown, … Susanne M Rafelski
    *Nature* (2023-01-04) https://doi.org/grkztd
    DOI: 10.1038/s41586-022-05563-7 · PMID: 36599983 · PMCID: PMC9834050

25. **Left-Ventricle Quantification Using Residual U-Net**
    Eric Kerfoot, James Clough, Ilkay Oksuz, Jack Lee, Andrew P King, Julia A Schnabel
    *Statistical Atlases and Computational Models of the Heart. Atrial Segmentation and LV Quantification Challenges* (2019) https://doi.org/gqdkdp
    DOI: 10.1007/978-3-030-12029-0_40

26. **Attention U-Net: Learning Where to Look for the Pancreas**
    Ozan Oktay, Jo Schlemper, Loic Le Folgoc, Matthew Lee, Matthias Heinrich, Kazunari Misawa, Kensaku Mori, Steven McDonagh, Nils Y Hammerla, Bernhard Kainz, … Daniel Rueckert
    *Proceedings of Medical Imaging with Deep Learning 2018* https://openreview.net/forum?id=Skft7cijM

27. **Swin UNETR: Swin Transformers for Semantic Segmentation of Brain Tumors in MRI Images**
    Ali Hatamizadeh, Vishwesh Nath, Yucheng Tang, Dong Yang, Holger R Roth, Daguang Xu
    *Brainlesion: Glioma, Multiple Sclerosis, Stroke and Traumatic Brain Injuries* (2022)
    https://doi.org/gqrg93
    DOI: 10.1007/978-3-031-08999-2_22

28. **UNETR: Transformers for 3D Medical Image Segmentation**
    Ali Hatamizadeh, Yucheng Tang, Vishwesh Nath, Dong Yang, Andriy Myronenko, Bennett Landman, Holger R Roth, Daguang Xu
    *2022 IEEE/CVF Winter Conference on Applications of Computer Vision (WACV)* (2022-01)
    https://doi.org/gqrg96
    DOI: 10.1109/wacv51458.2022.00181

29. **A Stochastic Polygons Model for Glandular Structures in Colon Histology Images**
    Korsuk Sirinukunwattana, David RJ Snead, Nasir M Rajpoot
    *IEEE Transactions on Medical Imaging* (2015-11) https://doi.org/gqrg95
    DOI: 10.1109/tmi.2015.2433900 · PMID: 25993703

30. **Gland segmentation in colon histology images: The glas challenge contest**
    Korsuk Sirinukunwattana, Josien PW Pluim, Hao Chen, Xiaojuan Qi, Pheng-Ann Heng, Yun Bo Guo, Li Yang Wang, Bogdan J Matuszewski, Elia Bruni, Urko Sanchez, … Nasir M Rajpoot
    *Medical Image Analysis* (2017-01) https://doi.org/c5t7
    DOI: 10.1016/j.media.2016.08.008 · PMID: 27614792

31. **A method for normalizing histology slides for quantitative analysis**
    Marc Macenko, Marc Niethammer, JS Marron, David Borland, John T Woosley, Xiaojun Guan, Charles Schmitt, Nancy E Thomas
    *2009 IEEE International Symposium on Biomedical Imaging: From Nano to Macro* (2009-06)
    https://doi.org/bmbj4h
    DOI: 10.1109/isbi.2009.5193250

32. **Cell Detection with Star-Convex Polygons**
    Uwe Schmidt, Martin Weigert, Coleman Broaddus, Gene Myers
    *Medical Image Computing and Computer Assisted Intervention – MICCAI 2018* (2018)
    https://doi.org/ggnzqb
    DOI: 10.1007/978-3-030-00934-2_30

33. **Star-convex Polyhedra for 3D Object Detection and Segmentation in Microscopy**
    Martin Weigert, Uwe Schmidt, Robert Haase, Ko Sugawara, Gene Myers
    *2020 IEEE Winter Conference on Applications of Computer Vision (WACV)* (2020-03)
    https://doi.org/gjp4g9
    DOI: 10.1109/wacv45572.2020.9093435

34. **Splinedist: Automated Cell Segmentation With Spline Curves**
    Soham Mandal, Virginie Uhlmann
    *2021 IEEE 18th International Symposium on Biomedical Imaging (ISBI)* (2021-04-13)
    https://doi.org/gqrg94
    DOI: 10.1109/isbi48211.2021.9433928

35. **Cellpose: a generalist algorithm for cellular segmentation**
    Carsen Stringer, Tim Wang, Michalis Michaelos, Marius Pachitariu
    *Nature Methods* (2020-12-14) https://doi.org/ghrgms
    DOI: 10.1038/s41592-020-01018-x · PMID: 33318659

36. **Omnipose: a high-precision morphology-independent solution for bacterial cell segmentation**
    Kevin J Cutler, Carsen Stringer, Teresa W Lo, Luca Rappez, Nicholas Stroustrup, S Brook Peterson, Paul A Wiggins, Joseph D Mougous
    *Nature Methods* (2022-10-17) https://doi.org/grnd95
    DOI: 10.1038/s41592-022-01639-4 · PMID: 36253643 · PMCID: PMC9636021

37. **Mask R-CNN**
    Kaiming He, Georgia Gkioxari, Piotr Dollar, Ross Girshick
    *IEEE Transactions on Pattern Analysis and Machine Intelligence* (2020-02-01)
    https://doi.org/gfxfwn
    DOI: 10.1109/tpami.2018.2844175 · PMID: 29994331

38. **ERFNet: Efficient Residual Factorized ConvNet for Real-Time Semantic Segmentation**
    Eduardo Romera, Jose M Alvarez, Luis M Bergasa, Roberto Arroyo
    *IEEE Transactions on Intelligent Transportation Systems* (2018-01) https://doi.org/gcs5h7
    DOI: 10.1109/tits.2017.2750080

39. **High-Throughput Screen for Novel Antimicrobials using a Whole Animal Infection Model**
    Terence I Moy, Annie L Conery, Jonah Larkins-Ford, Gang Wu, Ralph Mazitschek, Gabriele Casadei, Kim Lewis, Anne E Carpenter, Frederick M Ausubel
    *ACS Chemical Biology* (2009-06-29) https://doi.org/bdwdfc
    DOI: 10.1021/cb900084v · PMID: 19572548 · PMCID: PMC2745594

40. **Fully Unsupervised Probabilistic Noise2Void**
    Mangal Prakash, Manan Lalit, Pavel Tomancak, Alexander Krul, Florian Jug
    *2020 IEEE 17th International Symposium on Biomedical Imaging (ISBI)* (2020-04)
    https://doi.org/gjnw2v
    DOI: 10.1109/isbi45749.2020.9098612

41. **Interpretable Unsupervised Diversity Denoising and Artefact Removal**
    Mangal Prakash, Mauricio Delbracio, Peyman Milanfar, Florian Jug
    *Proceedings of the Tenth International Conference on Learning Representations*
    https://proceedings.mlr.press/v143/lalit21a.html

42. **Content-aware image restoration: pushing the limits of fluorescence microscopy**
    Martin Weigert, Uwe Schmidt, Tobias Boothe, Andreas Müller, Alexandr Dibrov, Akanksha Jain, Benjamin Wilhelm, Deborah Schmidt, Coleman Broaddus, Siân Culley, … Eugene W Myers
    *Nature Methods* (2018-11-26) https://doi.org/gfkkfd
    DOI: 10.1038/s41592-018-0216-7 · PMID: 30478326

43. **DeStripe: A Self2Self Spatio-Spectral Graph Neural Network with Unfolded Hessian for Stripe Artifact Removal in Light-Sheet Microscopy**
    Yu Liu, Kurt Weiss, Nassir Navab, Carsten Marr, Jan Huisken, Tingying Peng
    *Lecture Notes in Computer Science* (2022) https://doi.org/grxqnw
    DOI: 10.1007/978-3-031-16440-8_10

44. **A Diffusion Model Predicts 3D Shapes from 2D Microscopy Images**
    Dominik JE Waibel, Ernst Röell, Bastian Rieck, Raja Giryes, Carsten Marr
    *arXiv* (2022) https://doi.org/gqrthn
    DOI: 10.48550/arxiv.2208.14125

45. **napari: a multi-dimensional image viewer for Python**
    Nicholas Sofroniew, Talley Lambert, Kira Evans, Juan Nunez-Iglesias, Grzegorz Bokota, Philip Winston, Gonzalo Peña-Castellanos, Kevin Yamauchi, Matthias Bussonnier, Draga Doncila Pop, … Abigail McGovern
    *Zenodo* (2022-11-03) https://doi.org/gjpsxz
    DOI: 10.5281/zenodo.3555620